\documentstyle[11pt,paspconf,epsf]{article}

\begin{document}
\raisebox{0cm}[0cm][0cm]{\makebox[0cm][l]
{\hspace{-3.5cm}\parbox{16cm}{\em
To appear in "The Low Surface Brightness Universe", IAU Coll 171, eds. J.I.
Davies et al., A.S.P. Conference Series
}}}

\title{Found: High Surface Brightness Compact Galaxies}

\author{M.J. Drinkwater}
\affil{Physics, University of New South Wales, Sydney 2052, 
Australia}
\author{S. Phillipps, J.B. Jones}
\affil{Physics, University of Bristol, Tyndall Avenue, Bristol BS8 1TL, UK}
\author{M.D. Gregg}
\affil{IGPP, Lawrence Livermore Lab.,
L-413, Livermore, CA 94550, USA}
\author{Q.A. Parker}
\affil{Anglo-Australian Observatory, Coonabarabran, NSW 2357, Australia}
\author{R.M. Smith, J.I. Davies}
\affil{Physics \& Astronomy, University of Wales, Cardiff CF2 3YB, UK}
\author{E.M. Sadler}
\affil{School of Physics, University of Sydney, NSW 2006, Australia}

\begin{abstract}
We are using the 2dF spectrograph to make a survey of all objects
(``stars'' and ``galaxies'') in a 12 deg$^2$ region towards the Fornax
cluster. We have discovered a population of compact emission-line
galaxies unresolved on photographic sky survey plates and therefore
missing in most galaxy surveys based on such material. These
galaxies are as luminous as normal field galaxies. Using H$\alpha$ to
estimate star formation they contribute at least an additional
5\% to the local star formation rate.
\end{abstract}


\section{Introduction}

Most galaxy surveys only detect galaxies with a limited range of
surface brightness (SB): the difficulty of detecting low SB (LSB)
galaxies is well-accepted (Impey, Bothun \& Malin 1988, Ferguson \&
McGaugh 1995) but at the other extreme, it has been argued that there
is no selection against high SB (HSB) galaxies (Allen \& Shu 1979).
This was based on a small sample of bright galaxies, but the
conclusion has since been applied to nearly all galaxy samples based
on photographic sky surveys. Several groups have in fact detected
compact galaxies unresolved on photographic survey
plates, often termed compact emission-line galaxies
(CELGs).  Many have been found in optical QSO surveys (Guzman et al.\
1998, Stobie et al.\ 1997, Boyle et al.\ 1998).  Similar galaxies have
been found in the University of Michigan emission line galaxy survey
(Salzer et al.\ 1989) and among HII galaxies (Terlevich et
al.\ 1991). Unfortunately, none of these are from simple flux- or
size-limited samples so they cannot be used to determine the true
extent of the compact galaxy population.

In this paper we present the detection of a new population of compact
galaxies based on initial results of {\em The Fornax Spectroscopic
Survey} (see Drinkwater, Phillipps \& Jones, this volume), a complete
survey of {\em all} objects in a region of sky centred on the Fornax
cluster, being carried out with the 2dF multi-object spectrograph on
the AAT. Of 2947 ``stars'' and 1290 ``galaxies'' in the central 2dF
field down to $B_J=19.7$, we successfully observed 1249 (42\%) and
1250 (97\%) respectively. Among the ``stars'' we found a number of
objects with HII region-type spectra at redshifts of 10-50,000 km/s
making them field galaxies well beyond the Fornax cluster (1500 km/s).
These results are presented in more detail by Drinkwater et al.\ (1998).
 
\section{Properties of the New Compact Galaxies Galaxies}

\begin{figure}
\plotone{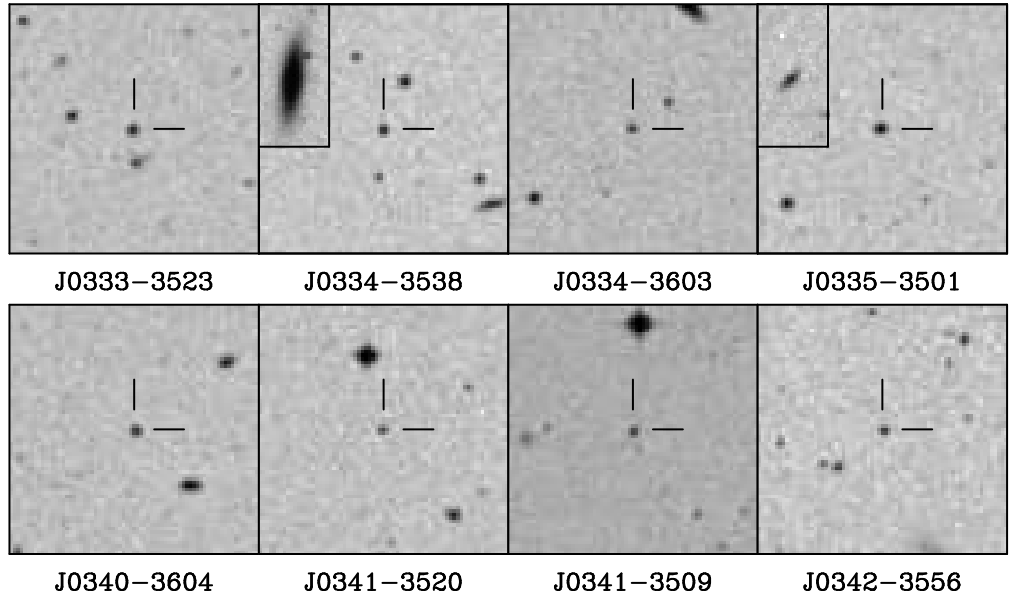}
\caption{$B_J$-band images of the new compact field
galaxies. Each is 2$'$ across. The insets in the images of
J0334$-$3538 and J0335$-$3501
show two normal ($M^\ast$) galaxies at the
same distances.
} \label{fig-images}
\end{figure}

We present images of the new compact galaxies in
Figure~\ref{fig-images}.
The new CELGs have similar absolute magnitudes and distances to
``normal'' field galaxies, but were classified as stars both by eye
and in the APM catalogue. For a typical seeing in survey plates (2")
this is equivalent to scale sizes less than about 1", or physical
scale sizes 0.6--2.2 kpc ($H_0=75$\,km\,s$^{-1}$\,Mpc$^{-1}$). These
upper limits are less than is typical of the local spiral galaxy
population (de Jong 1996).  However, these are not dwarfs in terms of
their luminosities which are within a factor 10 or so of L*; indeed
two of them exceed L*. This is clearly because of their high surface
brightnesses. If we take the limit of 1" to be the scale size of an
exponential profile, then the magnitudes imply central surface
brightnesses 19.5--21.8 $B$mag/arcsec$^2$, from 1 to 5 times brighter
than ``normal'' spirals or irregulars (van der Kruit 1987, de Jong
1996).  Interestingly, if we extrapolate to larger distances, the
CELGs have similar properties to the small galaxies in the Hubble Deep
Field. For instance, one of our CELGs placed at redshifts z=0.2--1
would have very similar scale lengths (0.1--0.2") and surface
brightness to those measured by Jones, Disney \& Phillipps (1998) for
disk-like galaxies in the HDF.

\section{Significance of the New Population}

The discovery of these CELGs poses several questions, notably what
fraction of all galaxies has been missed?  At redshifts less than
55000 km/s we observed 600 resolved objects (galaxies): about 300 of
these exhibited strong emission line features.  Within the same
distance and magnitude limits we found seven new CELGs. These
therefore constitute about 1\% of the observed local galaxy population, but
have been missed by existing surveys. All the new compacts have strong
H$\alpha$ emission, so they represent twice this fraction of emission
line galaxies (2\%). We can go one step further and consider their
contribution to the local star formation rate as H$\alpha$ emission
is a measure of recent star formation (Tresse \& Maddox 1998). Our
spectra are not flux-calibrated and the 2" apertures only partially
sample the flux from resolved objects. However the equivalent width of
the emission lines is proportional to the absolute flux in the
emission lines of the different galaxies to first order if we assume
that the galaxies all come from the same distribution of intrinsic
magnitudes and distances as shown above. On this basis the seven new
compact galaxies contribute an extra 5\% of star formation rate over
that in the whole sample of 600. As we have surveyed less than half of
the unresolved objects, these numbers are likely to double in the
final analysis.

%

\end{document}